\documentclass[prb,showpacs,twocolumn,superscriptaddress]{revtex4}

\usepackage{amsmath,graphicx,latexsym,bm}

\begin{document}

\title{From flexoelectricity to absolute deformation potentials: The case of SrTiO$_3$}

\author{Massimiliano Stengel}
\affiliation{ICREA - Instituci\'o Catalana de Recerca i Estudis Avan\c{c}ats, 08010 Barcelona, Spain}
\affiliation{Institut de Ci\`encia de Materials de Barcelona 
(ICMAB-CSIC), Campus UAB, 08193 Bellaterra, Spain}

\date{\today}

\begin{abstract} 
Based on recent developments in the first-principles theory of flexoelectricity, 
we generalize the concept of \emph{absolute deformation potential}
to arbitrary nonpiezoelectric insulators and deformation fields.
To demonstrate our formalism, we calculate the response of
the band edges of SrTiO$_3$ to both dynamic (sound waves) and static 
(bending) mechanical loads, respectively at the bulk level and in a slab 
geometry.
Our results have important implications for the understanding of strain-gradient-related
phenomena in crystalline insulators, formally unifying the description of band-structure
and electrostatic effects.

\end{abstract}

\pacs{71.15.-m, 
       77.65.-j, 
        63.20.dk} 
\maketitle

\section{Introduction}

The so-called deformation potentials, introduced in 1950 
by Bardeen and Shockley,~\cite{bardeen-50} 
describe the shifts in the single-particle electronic 
energy levels that are induced by a strain field.
Originally aimed at estimating the intrinsic contributions
to carrier mobility in semiconductors, this concept is nowadays important 
for a wide range of phenomena of fundamental and technological interest.
For example, deformation potentials were invoked to explain 
band offset formation at strained semiconductor 
heterojunctions~\cite{Peressi-98},
carrier confinement in optoelectronic 
devices,~\cite{Janotti-07} and the impact of dopants and 
impurities on the lattice parameter of insulators~\cite{Bruneval:2015,
Janotti-12}.
In all the above contexts, one needs to know how the energy of
a given electronic state is modulated by an inhomogeneous strain
field, which can be produced by an acoustic phonon, by a 
compositional gradient or by an applied external load.
This physical property goes now
by the name of \emph{absolute deformation potential} (ADP), where the
qualification ``absolute'' was introduced~\cite{vandewalle-89}
to emphasize the focus on the single energy levels, rather than 
their relative differences.

For theoretical analysis and computational purposes, ADPs are typically 
split into two separate contributions, a macroscopic electrostatic (ME) term 
and a band-structure (BS) term. 
The latter consists in the strain derivative of the Bloch eigenvalues,
where the (arbitrary) energy reference is fixed to the average electrostatic
potential (or to the Fermi level in a metal).
This is manifestly a bulk property, as it can be readily obtained by
performing periodic calculations that involve the primitive unit cell of
the crystal only.
The former, which is by far the most challenging to calculate, 
consists in the strain derivative of the electrostatic reference itself.
(This term is due to long-range electrostatic interactions, and
therefore it is only present in insulators.)
As such reference is generally ill-defined in an infinite crystal,~\cite{Kleinman-81}
one needs to explicitly consider an inhomogeneous strain field, where the 
cell parameters gradually evolve between two different configurations A and B.
The integral of the electric field along the AB path yields then the sought-after
voltage lineup.
The nonanaliticity of electrostatic interactions implies that the lineup
depends not only on the end points, A and B, but also on the specific path
connecting them.
Yet, once the direction of such a path is known (assuming it is rectilinear),
this term can also be expressed as a bulk property,~\cite{resta-dp,resta-prb91} 
and explicit first-principles calculations have been reported for a number of 
materials.~\cite{vandewalle-89,resta-dp,Janotti-07}

In all the aforementioned works, the focus has been mainly restricted to a 
particular class of deformations (longitudinal acoustic phonons) and dielectrics
(nonpolar semiconductors like Si or Ge). These essentially reflect the
established range of validity of the current ADP theory.~\cite{resta-dp}
There are increasingly good reasons, however, to seek a more general 
approach to the problem in order to overcome the above limitations.
Regarding the materials issue, there has been recently increasing interest
in semiconducting oxides: ZnO, SrTiO$_3$, etc. with highly confined
2D electron gases. Acoustic phonon scattering has been identified as
one of the main factors limiting mobility, e.g., at the LaAlO$_3$/SrTiO$_3$ 
interface, either in the superconducting state~\cite{Klimin-14}, or in
thermoelectric applications~\cite{Pallecchi-15}; in other 
cases, ADPs were invoked to explain the carrier confinement mechanism
itself.~\cite{Janotti-07}
Polar materials (in this work we shall indicate as ``polar'' all crystals 
where the Born charge tensor does not vanish) presently fall outside the 
scopes of ADP theory,~\cite{resta-dp} and this poses a clear problem for the 
interpretation of the above phenomena.
Regarding the ``geometric'' issue, there has been recently a strong interest 
in the electronic properties of bent nanostructures, and in particular
in determining how curvature affects the effective potential felt by
quantum particles.~\cite{ortix} Accurate knowledge of the bending-related 
ADPs is necessary to obtain a quantitative solution to this problem,~\cite{ortix}
but bending is a type of deformation that cannot be described as a 
longitudinal acoustic phonon; this seriously limits the applicability of
first-principles methods in this context.

Here we show that the theory of flexoelectricity, which has undergone 
an impressive development in the past few years, is now mature enough 
to provide a rigorous answer to both questions. 
Flexoelectricity (FxE) describes the electrical polarization that is linearly
induced by a strain gradient, and is therefore ideally suited to 
tackle the macroscopic electrostatic contribution to the ADPs in the
most general case.
(Indeed, the longitudinal components of the macroscopic FxE tensor 
reduce~\cite{Resta-10} to the ME term of Ref.~\onlinecite{resta-dp} in 
a nonpolar crystal.)
In particular, we argue that (i) the atomic relaxations (internal strains) that occur 
in a strain-gradient field are bulk properties of the material
(just like the purely electronic effects that were considered in earlier 
works~\cite{resta-prb91,resta-dp}), 
thereby extending the ADP theory to polar insulators; and (ii) the 
\emph{transverse} components of the FxE tensor naturally yield the
desired information on the bending-induced electrostatic effects, which 
allows us to formally extend the ADP theory to arbitrary deformation fields.
As a practical demonstration, we provide a full calculation of the ADPs 
in cubic SrTiO$_3$, considering both the dynamic regime of sound waves and
the static one of a bent slab. 

Apart from the immediate relevance to the physics of SrTiO$_3$, these
results highlight a crucial aspect of flexoelectricity that is absent
in other types of electromechanical couplings: As each energy level 
experiences a different tilt, in a macroscopic strain gradient the 
very notion of \emph{macroscopic} electric field is inherently ambiguous, 
and depends on the physical nature (electron, hole, classical ``test'' charge, etc.) 
of the charged particle that is used to define it.
An elegant solution to this conceptual puzzle consists in 
\emph{identifying} the flexovoltage coefficients with the ADPs,
and thereby rationalizing the aforementioned ambiguity via the band-energy
dependence of the deformation potential.
This points to a formal unification of ADP and flexoelectricity
theories, with important implications for both the interpretation
of the experiments and the theoretical modeling of strain-gradient 
related phenomena.

\section{Theory}

Consider a macroscopic deformation where a given component of the
symmetric stress tensor, $\varepsilon_{\beta \gamma}$, undergoes a 
linear increase along the direction $\hat{\bf q}$. The absolute deformation potential
of the $n$-th band at the point ${\bf k}$ in the Brillouin zone can be written as
\begin{equation}
D^{\beta \gamma, \hat{\bf q}}_{n \kappa} = \frac{ d E_{n {\bf k}} }{ d \varepsilon_{\beta \gamma}}
 - \frac{e}{\epsilon_0 \epsilon_{\hat{\bf q}}} \mu^{\rm II}_{\alpha \lambda, \beta \gamma} \, \hat{q}_\alpha \hat{q}_\lambda.
\label{adp}
\end{equation} 
Here $E_{n {\bf k}}$ is the energy eigenvalue referred to the mean electron potential 
energy, 
\begin{equation}
\mu^{\rm II}_{\alpha \lambda, \beta \gamma} = \frac{d P_\alpha}{d \varepsilon_{\beta \gamma,\lambda}}
\end{equation}
is the type-II flexoelectric tensor (referring to the derivative of the polarization
with respect to the strain gradient component $\varepsilon_{\beta \gamma,\lambda}$,
where the latter is the gradient of $\varepsilon_{\beta \gamma}$ along the Cartesian
direction $r_\lambda$), $\epsilon_{\hat{\bf q}} = \hat{\bf q} \cdot \bm{\epsilon} \cdot \hat{\bf q}$
is the relative permittivity along $\hat{\bf q}$, and $\epsilon_0$ is the vacuum permittivity.
The first term on the rhs of Eq.~(\ref{adp}) is the band-structure term, and is 
independent of the direction $\hat{\bf q}$; following earlier works,~\cite{vandewalle-89,resta-dp}
we shall focus on the valence ($v$) and conduction ($c$) band edges, and indicate the corresponding
BS terms as $\Delta D_{v,c}^{\beta \gamma}$.
The second term in Eq.~(\ref{adp}) is \emph{minus} the electric field
that is generated by the strain gradient $\varepsilon_{\beta \gamma,\lambda} \hat{q}_\lambda$
when open-circuit boundary conditions are imposed along $\hat{\bf q}$; this corresponds
precisely to the macroscopic electrostatic term in the ADP. (Note the $-e$ factor, reflecting the
negative electronic charge.)

To see that the present theory recovers earlier treatments~\cite{resta-dp} as 
special cases, consider the macroscopic displacement pattern associated with an acoustic 
phonon of wavevector ${\bf q}$,
\begin{equation}
u^l_{\beta \kappa} = U_\beta e^{i{\bf q} \cdot {\bf r}},
\end{equation}
where $l$ is a cell index, $\kappa$ is a sublattice index, and $U_\beta$
is a real-space vector indicating the displacement amplitude and direction.
For a longitudinal phonon, the symmetric strain is
\begin{equation}
\varepsilon_{\beta \gamma} ({\bf r}) = i qU \hat{q}_\beta \hat{q}_\gamma e^{i{\bf q} \cdot {\bf r}},
\end{equation}
where $q=|{\bf q}|$ and $U=|{\bf U}|$.
We readily obtain, for the ME part,
\begin{equation}
D^{\hat{\bf q}}_{\rm (macro)} = 
 - \frac{e}{\epsilon_0 \epsilon_{\hat{\bf q}}} \mu_{\hat{\bf q}}, \qquad
 \mu_{\hat{\bf q}} = \mu^{\rm II}_{\alpha \lambda, \beta \gamma} \, \hat{q}_\alpha \hat{q}_\beta \hat{q}_\gamma \hat{q}_\lambda.
\end{equation}
It is then straightforward to show~\cite{artlin} that, in the case of a nonpolar insulator,
$D^{\hat{\bf q}}_{\rm (macro)}$ reduces to the expression of Ref.~\onlinecite{resta-dp},
written in terms of the dynamical quadrupole and octupole tensors.
In a material with cubic symmetry this can be written, in turn, as~\cite{resta-dp}
\begin{equation}
D_{\rm macro}(\hat{\bf q}) = A_{\rm macro} + f(\hat{\bf q}) B,
\label{amacro}
\end{equation}
where the anisotropy function 
$f(\hat{\bf q})= 3 (1 - \sum_\alpha \hat{q}^4_\alpha) / 2$
varies between 0 and 1 depending on the direction.
Then, the final result is
\begin{equation}
D_{v,c}(\hat{\bf q}) = A_{v,c} + f(\hat{\bf q}) B, 
\end{equation}
where $A_{v,c} = A_{\rm macro} + \Delta D_{v,c}$.

Having shown the consistency of Eq.~(\ref{adp}) with earlier works,
we can now move on to clarifying in what respects such expression 
generalizes the scopes of the ADP theory to a broader range of
phenomena. 
In a nutshell, by using Eq.~(\ref{adp}) to define the ADPs,
one can in principle tackle all
materials (including polar crystals) and deformation fields (including bending)
for which the flexoelectric tensor is well defined.
To appreciate the practical implications of this statement, it is useful to discuss 
in some detail the example of a polar (but nonpiezoelectric~\cite{artlin}) insulator.
In Ref.~\onlinecite{resta-dp}, the authors argued that ADPs cannot be written as
bulk material properties in this case, as the displacement of a plane of 
atoms induces a net dipole density, and hence an arbitrary shift in the electrostatic 
lineup. 
The theory of flexoelectricity, however, shows that such displacements are \emph{not}
arbitrary, but can be themselves written as bulk material properties, via a higher-order
internal-strain tensor (indicated as ${\bf N}$ or ${\bf L}$ in Ref.~\onlinecite{artlin}).
This allows one to write, in full generality, the \emph{lattice-mediated} (LM) contribution to the 
ADPs, which for a LA phonon in a cubic insulator has the same form as Eq.~(\ref{amacro}).
(The frozen-ion and relaxed-ion ADPs retain the same functional form, only with different
values of the coefficients $A_{\rm (macro)}$ and $B$.)

It is worth making, in this context, an additional comment on the physical 
nature of the LM contribution. 
Such contribution depends, as we said, on the flexoelectric internal-strain tensor components,
and this is an inherently \emph{dynamic} quantity, i.e. it depends explicitly on the atomic masses;~\cite{artlin}
this characteristic obviously propagates to the relaxed-ion ADPs.
This is not a concern when studying, e.g., the impact of acoustic vibrations 
(a dynamic effect) on carrier mobility, but questions, at first sight, the applicability
of the present theory to genuinely static cases, such as a bent slab under an external
mechanical load.
Note, however, that the condition of static equilibrium imposes a precise linear
relationship between the strain-gradient components that are present in a given region
of the sample interior. (For example, in a bent plate the ``primary'' transverse deformation is
always accompanied by a ``secondary'' longitudinal strain gradient that is oriented along
the surface normal.~\cite{artgr,artcalc})
Once these elastic relaxation effects are appropriately incorporated, the mass-dependent terms that are 
present in the internal-strain tensor ${\bf L}$ exactly cancel out.~\cite{artlin} This clears
up the above worries regarding lattice-mediated effects: the ADPs described by Eq.~(\ref{adp})
are indeed \emph{static} quantities in the slab-bending case, while they are \emph{dynamic} 
in the context of LA phonons, consistent with the physical nature of the phenomenon under study.

\section{Results}

\subsection{Longitudinal acoustic phonons}

\begin{figure}
\begin{center}
\includegraphics[width=3.3in]{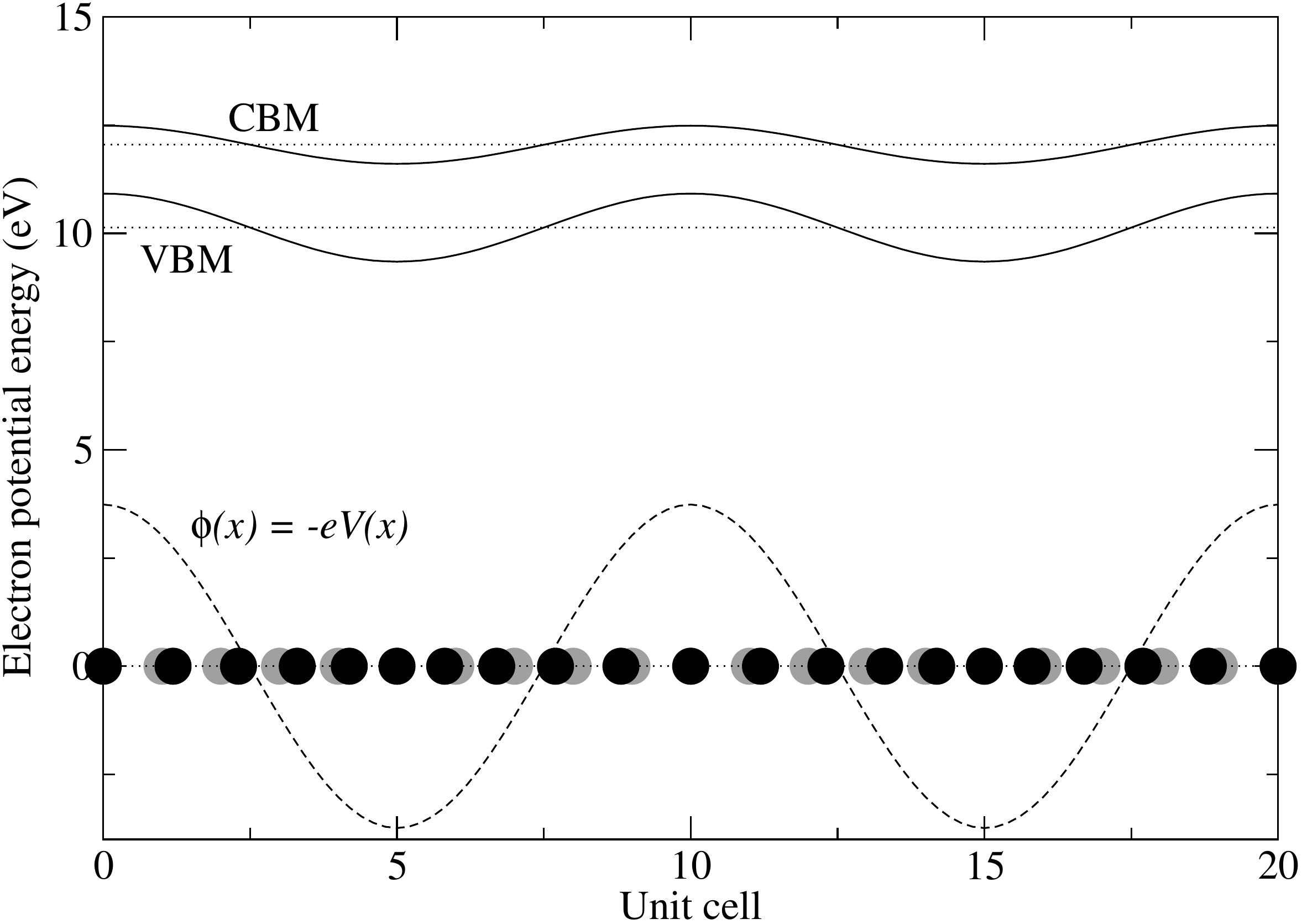}
\end{center}
\caption{ \label{fig1} Illustrative sketch of the deformation potentials in a [100]-oriented 
sound wave in SrTiO$_3$. Conduction (CBM) and valence (VBM) band edges are shown, as well as
the mean electron potential energy, $\phi(x) = -eV(x)$. ($V$ stands for the electrostatic
potential.)
The snapshot of the deformation is shown as black circles displacing from
their equilibrium position (gray circles). An unrealistically large local strain of
$\pm20$\% (at the extremes of the sinusoidal wave) is used for illustration purposes.}
\end{figure}

\begin{table}
\begin{center}
\begin{ruledtabular}
\begin{tabular}{crrrr}
 & $A_{\rm macro}$  & $A_{c}$  & $A_{v}$ & $B$ \\
\hline 
 Frozen-ion &   16.15  & $-$0.33 & 1.40 & 1.29 \\
 Total      &   18.68  &    2.20 & 3.93 & 0.98                               
\end{tabular}
\end{ruledtabular}
\end{center}
\caption{ Deformation potentials referring to longitudinal phonons along an 
arbitrary direction. Values are in eV. \label{tab1}}
\end{table}

In the following, we shall illustrate the above arguments by explicitly calculating the
relaxed-ion ADPs in SrTiO$_3$.
Our calculations are performed within the local-density approximation~\cite{Perdew/Wang:1992} 
to density-functional theory. We use the same pseudopotentials and computational parameters as 
in Ref.~\onlinecite{artcalc}, from which we also take the numerical values of the relevant
bulk flexoelectric tensor components. These data are then processed to yield the macroscopic
electrostatic contributions to the ADPs. The band-structure terms, $\Delta D_{v,c}^{\beta \gamma}$
are calculated by finite differences, considering an isotropic volume expansion (or compression) 
of the cubic SrTiO$_3$ cell. This corresponds to neglecting the splittings induced by anisotropic
deformations, consistent with earlier works.~\cite{vandewalle-89,resta-dp} 
The reported values correspond to the valence-band top at the $R$ point of
the Brillouin zone, and to the conduction-band bottom at $\Gamma$.

We shall start by discussing our results for the ``acoustic'' ADPs, 
which are described by Eq.~(\ref{amacro}).
Our calculated numerical values for $A_{\rm macro}$, $A_v$, $A_c$ and $B$,
both at the frozen-ion and relaxed-ion levels, are reported
in Table~\ref{tab1}.
The case of a [100]-oriented LA phonon is also schematically illustrated in Fig.~\ref{fig1}.

First, note the large difference in absolute value between $A_{\rm macro}$ and $A_{v,c}$, 
which points to a substantial
cancellation between the (positive) macroscopic and (negative) band-structure 
terms. A closer look at the numerical data indicates that such cancellation 
occurs at the purely electronic (frozen-ion) level.
To rationalize this outcome (which is a common feature of ADP 
calculations~\cite{resta-dp}), 
recall that the frozen-ion value of 
the longitudinal flexoelectric coefficients (and hence of $A_{\rm macro}$)
are related~\cite{Hong-13,artlin} to the octupolar moments of the electronic charge 
that each atom ``drags'' along during the deformation. 
Such moments are typically dominated by a spherical term, which originates
from the inner orbitals that rigidly follow each nucleus 
along its distortion path.
These orbitals, in turn, are chemically inert, meaning that their displacement
contribute to $A_{\rm macro}$ but not to the band energies.
(In a ideally ionic solid, e.g. like the noninteracting rigid-ion model of Ref.~\onlinecite{artgr},
the cancellation would be complete, and yield
vanishing values for $A_{v,c}$.)

Next, it is interesting to note the relatively large and positive 
contribution from lattice-mediated effects. (Once relaxation effects are
accounted for, both bands undergo an \emph{upward} shift upon uniaxial
tension, see Fig.~\ref{fig1}.)
This observation can be readily explained by recalling the dynamical 
nature of the lattice-mediated flexoelectric effect in sound waves.
Indeed, a strain gradient dynamically induces (in addition to
the mass-independent contributions from the interatomic force constants) 
a force on each ionic sublattice that is directly proportional to its nuclear
mass~\cite{artlin}.
Since the cations (Sr, Ti) are much heavier than the anions (O), this
produces a systematic bias on the sign of the flexoelectric coefficients,
consistent with the results of Table~\ref{tab1}.
To verify such a hypothesis, we have performed a computational experiment,
where we have recalculated the lattice-mediated contribution to $A_{\rm macro}$,
$A_{\rm macro}^{\rm LM}$, while setting all nuclear masses to a uniform value.
As a result, $A_{\rm macro}^{\rm LM}$ changed from 2.53 eV to $-$1.88 eV, confirming
our arguments. (A closely related analysis was also performed in 
Ref.~\onlinecite{Hong-13}, with qualitatively similar conclusions, see Table IV 
therein.)

Finally, note the relatively small anisotropy, which is only marginally affected
by lattice-mediated effects, and the comparably larger value of $A_v$ respect to
$A_c$. The latter observation indicates that the gap of SrTiO$_3$ shrinks upon
volume expansion, consistent with the dominant ionic character of the bonding.
(Our calculated gap deformation potential is $A_{\rm gap}=-1.73$ eV.)

Our frozen-ion results can be directly compared with those that were recently 
reported by Janotti {\em et al.}~\cite{Janotti-12}
There appears to be a significant discrepancy ($A_v=-1.5$ eV, $A_c=-4.0$ eV,
$A_{\rm gap}=-2.5$ eV in Ref.~\onlinecite{Janotti-12}), whose precise origin 
is unclear at present. 
One possible source of disagreement may stem from the different exchange and
correlation functional that was used therein (HSE).~\cite{hse}
This, however, would imply a worrisome dependence of the ADPs on the
specific details of the computational model; further investigations will be 
necessary in order to clarify this important point.

\subsection{Static bending of a (100)-oriented plate}

\begin{figure}
\begin{center}
\includegraphics[width=3.3in]{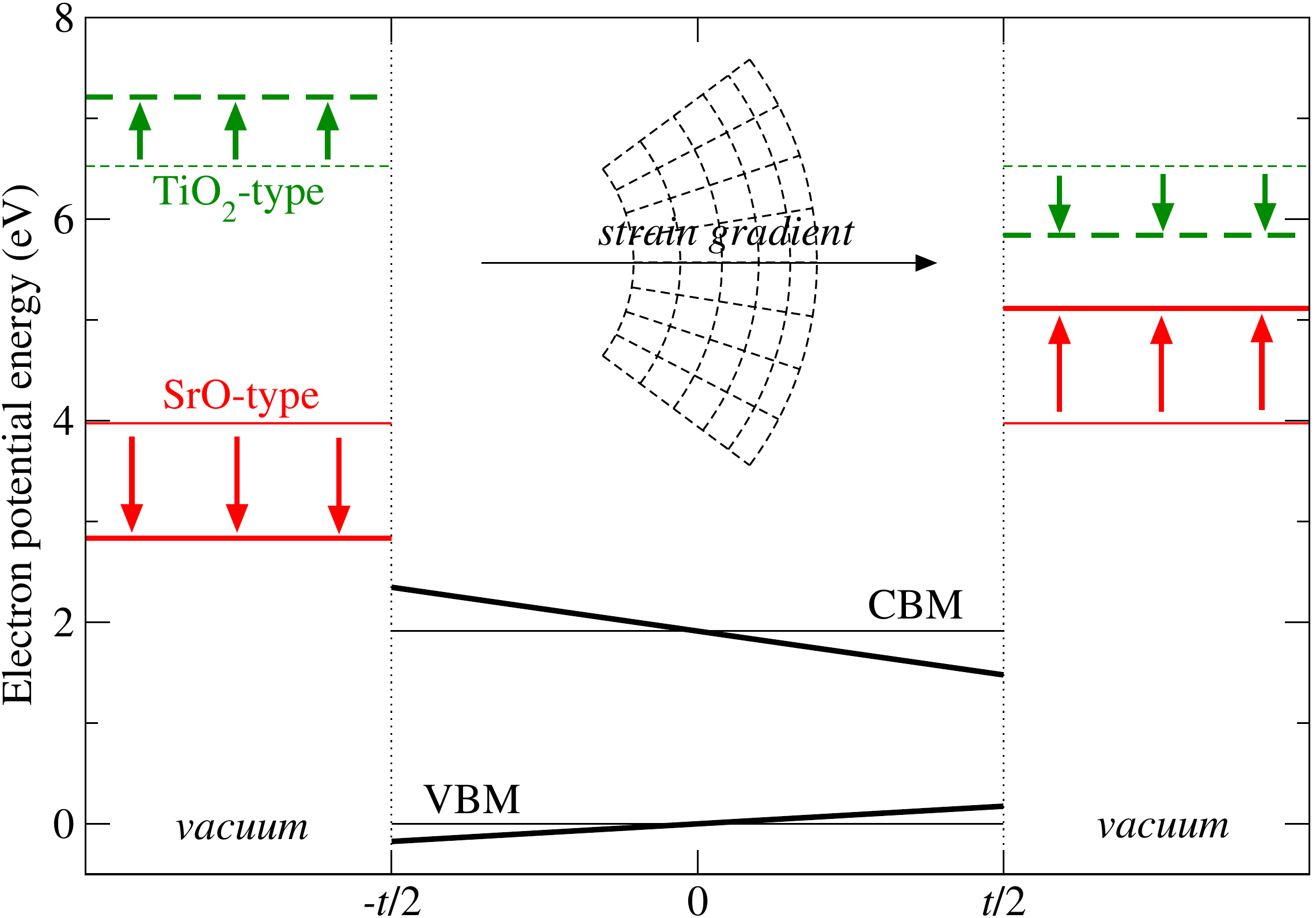}
\end{center}
\caption{ \label{fig2} Sketch of the relevant energy levels in a bent SrTiO$_3$ slab.
Thin solid black lines indicate the flat conduction and valence band edges in the unperturbed 
system. Thin dashed black lines indicate the slab surfaces. Colored thin lines indicate the
vacuum level. Upon bending, the band edges tilt and the vacuum levels shift (thick lines). An 
unrealistically large strain gradient, corresponding to $t \cdot \varepsilon_{yy,x} =1$ is assumed
for illustrative purposes. The inset illustrates the type (not the amplitude) of the deformation
applied to the slab.
}
\end{figure}

\begin{table}
\begin{center}
\begin{ruledtabular}
\begin{tabular}{crrr}
 & $\varphi_{\rm macro}$  & $\varphi_{c}$  & $\varphi_{v}$ \\
\hline 
 Frozen-ion$^{(*)}$ &   10.37  & $-$1.31 & $-$0.08 \\
 Total              &   10.81  & $-$0.87 &    0.36                               
\end{tabular}
\end{ruledtabular}
\end{center}
\caption{ Deformation potentials associated 
with static bending of a SrTiO$_3$ slab (plate-bending limit).
The surface is assumed to be oriented along the [100] cubic direction.
The (*) symbol indicates that ``frozen-ion'' here is referred to
a linear combination of transverse and longitudinal strain gradients,
consistent with the mechanical equilibrium conditions. Values are
in eV. \label{tab2}}
\end{table}

Our flexoelectricity-based theory of ADPs can readily answer another 
important physical question~\cite{ortix} that was so far unaddressed at the first-principles level.
This concerns the modifications to the electronic structure of a slab that are induced by a \emph{bending}
deformation.
The validity of the ADP concept in this specific case rests on the demonstration, provided
in Ref.~\onlinecite{artgr}, that the open-circuit internal field of a slab subjected to bending is 
a bulk material property.
As we mentioned earlier, it is most appropriate here to assume mechanical equilibrium 
(under the action of an external load), and therefore consider \emph{static} ADPs.
These can be expressed as effective linear combination of the transverse and longitudinal flexovoltage 
coefficients, and decomposed according to Eq.~(\ref{adp}) into a macroscopic and a band-structure term,
\begin{equation}
D^{\rm eff}_{v,c} = D^{\rm eff}_{\rm macro} + \Delta D^{\rm eff}_{v,c}.
\end{equation}
Here $D^{\rm eff}_{\rm macro}=-e \varphi_{\rm eff}$ corresponds to \emph{minus} the effective 
flexovoltage coefficient,~\cite{artcalc} $\varphi_{\rm eff}$, scaled by the electron charge.
The band-structure term, on the other hand, consists in the usual isotropic volume shift with a
prefactor,~\cite{artcalc} $1-\nu$, that accounts for Poisson's ratio effects, 
\begin{equation}
D^{\rm eff}_{\rm BS} = (1-\nu) \frac{d E_{v,c}}{d \ln (\Omega)}.
\end{equation}
($\nu=c_{12}/c_{11}$, where $c_{ij}$ is the bulk elastic tensor in Voigt notation.)

As a testcase, we considered a (100)-oriented slab of SrTiO$_3$ in the plate-bending 
regime, analogous to Ref.~\onlinecite{artcalc}. 
(A more extensive analysis, involving slabs of different orientations and 
a comparison of the beam-bending versus plate-bending limits, is reported
in Section~\ref{sec:flexo}.)
In Table~\ref{tab2} we show our results 
for $D^{\rm eff}_{\rm macro}$ and $D^{\rm eff}_{v,c}$. Again, we calculated both quantities 
at the frozen-ion level first (note that here ``frozen-ion'' implies neglect of the internal strains,
but inclusion of the aforementioned Poisson's ratio effects), and later incorporating full atomic
relaxation.
As in the phonon case discussed earlier, there is a large cancellation
between the macroscopic and the band-structure terms at the frozen-ion level.
Due to this cancellation, the relative impact of the lattice-mediated contribution,
small on the scale of the large macroscopic term,
becomes significant in the context of the ADPs (the latter are about an order 
of magnitude smaller than $D^{\rm eff}_{\rm macro}$); for example,
the valence-band parameter $D^{\rm eff}_v$ changes sign after relaxation. 
Compared to the phonon ADPs of Table~\ref{tab1}, here the relaxed-ion 
values are significantly smaller in magnitude; also, the conduction-band
and valence-band ADP have opposite signs. 
This corroborates our interpretation that the large and positive values
of the phonon ADPs  are mainly due to dynamical effects,
which are absent in the present slab-bending case.
The above results are schematically illustrated in Fig.~\ref{fig1}, where we also have incorporated
the results of Ref.~\onlinecite{artcalc} regarding the surface contributions to the open-circuit 
flexovoltage.

\subsection{Back to flexoelectricity}
\label{sec:flexo}

The energy level diagram of Fig.~\ref{fig1} may appear, at first sight, confusing:
in the slab interior we have represented band-structure effects (which are the
realm of ADP theory), while in the vacuum regions we have plotted the electromechanical
response coefficients of Ref.~\onlinecite{artcalc} (which were derived in the 
context of flexoelectricity).
Such a juxtaposition was made on purpose -- as we shall explain shortly, 
there is an even more intimate link between flexoelectricity and
ADPs than the earlier sections of this work suggest.

To see this, consider the total open-circuit voltage induced by bending, $\varphi^{\rm total}$,
which in Fig.~\ref{fig2} corresponds to (minus) the net shift between the (constant) 
vacuum potentials located at either side of the bent slab. (This is
the physical quantity that flexoelectric experiments typically focus on.)
In Ref.~\onlinecite{artcalc}, such a quantity was obtained by summing up the 
relevant bulk flexovoltage coefficient (corresponding to $\varphi^{\rm bulk}$ 
i.e. the tilt of the macroscopic electrostatic potential in the slab interior) 
and the surface dipolar contribution, $\varphi^{\rm surf}$,
\begin{equation}
\varphi^{\rm total} = \varphi^{\rm bulk} + \varphi^{\rm surf}.
\end{equation}
Fig.~\ref{fig2} shows that the same quantity can be, equivalently, obtained
by summing up the bulk ADP (either the valence-band $D^{\rm eff}_v$ or 
the conduction-band $D^{\rm eff}_c$), and the appropriate \emph{surface
deformation potential}~\cite{Zhou-13}, $D^{\rm surf}_{v,c}$,
\begin{equation}
-e \varphi^{\rm total} = D^{\rm eff}_{v,c} + D^{\rm surf}_{v,c}.
\end{equation}
(The latter is the strain derivative of the offset between the conduction or valence 
band edges and the vacuum level; such offsets are commonly known in the literature 
by the name of electron affinity and ionization potential, respectively.)
In our slab models of SrTiO$_3$, we find $D^{\rm SrO}_{v}=1.92$ eV,
and $D^{\rm TiO_2}_{v}=-1.73$ eV, where the superscript refers to the two types of 
surface terminations considered in Ref.~\onlinecite{artcalc}.
(The corresponding values for the conduction band surface deformation potentials 
can be trivially determined by using the data of Table~\ref{tab2}.)

The fact that, by summing up either the bulk and surface flexovoltage 
coefficients or the corresponding valence-band ADPs, we have obtained the same number
(the total flexovoltage response of the slab) is not fortuitous.
Flexovoltage coefficients really are ADPs in disguise (they can be thought
as ``electrostatic potential ADPs'')  and, similarly, ADPs can be perfectly 
well regarded as flexovoltage coefficients.
The former statement is an obvious consequence of Eq.~(\ref{adp}): by setting
the band-structure term to zero, and by recalling the relationship between
flexovoltage and flexoelectric coefficients~\cite{artcalc}, one trivially 
obtains $\varphi^{\rm bulk} = - D^{\rm eff}_{\rm macro}/e$.
Accepting the latter statement, on the other hand, may seem awkward at first sight, 
as it implies that the bulk flexoelectric tensor is not uniquely defined. 
(ADPs depend on the specific band feature that is being considered.)
This is, however, perfectly consistent with the fundamental theory of 
flexoelectricity, where such a nonuniqueness emerges as a natural consequence 
of the formalism.~\cite{artlin} 

To appreciate the physical origin of this arbitrariness, one does not 
necessarily need to go through pages of complicated algebra.
In fact, it suffices to take a closer look at Fig.~\ref{fig2} and
Table~\ref{tab2} to get a reasonably clear idea of what's going on: 
The valence band, conduction band and mean electrostatic potential 
all undergo a different tilt. 
This means that a conduction electron, a valence hole and a classical 
``test'' charge will not feel the same electrical force. Otherwise stated, 
in a uniform strain gradient the very definition of ``macroscopic electric field'', ${\bf E}$, 
\emph{depends} on the physical nature of the charged particle that is used to probe it.
As the ${\bf E}$-field in the interior of a bent slab is, in the above sense, 
ambiguous, the notion of ``short-circuit boundary conditions'' (a necessary 
ingredient for defining the bulk flexoelectric tensor) is equally ambiguous therein -- 
it depends on the (arbitrary) energy reference that we choose when imposing the ${\bf E}=0$ 
(i.e. flat-band) condition.~\footnote{Of course, one can always think of depositing metal 
  electrodes onto the slab surfaces -- this way,
  the short-circuit condition is no longer arbitrary. Still, the precise way that a ``zero 
  external potential'' translates into a ``vanishing internal field'' depends on the properties 
  (i.e. on the interface deformation potential) of the film-electrode interface. In some sense,
  the interface/surface ``selects'' a given energy reference out of the infinite possibilities 
  that exist at the bulk level, thereby lifting the aforementioned arbitrariness.}

Now, while it is true that selecting one or the other reference potential to define the internal 
electric field of the slab does not affect the overall result (the total open-circuit
voltage), this choice does modify the way the effect is split into surface and bulk 
contributions.
In practice, such a freedom can be exploited to achieve a more meaningful 
physical description of the two individual pieces.
In the present case of the SrTiO$_3$ slab, for example, it appears
tempting to identify the relevant flexovoltage coefficients with the valence-band 
(VB) ADPs,
\begin{equation}
\varphi^{\rm bulk-VB} = -\frac{D^{\rm eff}_v}{e}, \qquad \varphi^{\rm surf-VB} = -\frac{D^{\rm surf}_v}{e},
\end{equation}
rather than with the gradient of the bare electrostatic potential as it was
done earlier~\cite{artcalc}.
This way, we obtain a partition between bulk and surface effects where both 
contributions are small in magnitude (i.e. of the
same order as the total open-circuit flexovoltage of the slab), thereby
facilitating the identification of the physical mechanisms that
play a dominant role in either context.
(Recall that the ``electrostatic'' flexovoltages are typically characterized 
by a large cancellation between bulk and surface contributions,~\cite{artcalc} 
which complicates the physical interpretation of the two individual terms.)

\subsection{Bending anisotropy of SrTiO$_3$ slabs} 

\begin{table}[!t]
\begin{center}
\begin{ruledtabular}
\begin{tabular}{c rrr}
 \multicolumn{4}{c}{Conduction band} \\
    &  (100) & (110) &  (111)  \\
\hline 
 Plate & $-$0.87 & $-$1.97 & $-$1.78 \\
 Beam  & $-$0.68 & $-$1.73 & $-$1.33 \\

 \multicolumn{4}{c}{ } \\
 \multicolumn{4}{c}{Valence band} \\
    &  (100) & (110) &  (111)  \\
\hline 
 Plate & 0.35 & $-$0.83 & $-$0.63 \\
 Beam  & 0.27 & $-$0.85 & $-$0.47 
\end{tabular}
\end{ruledtabular}

\end{center}
\caption{ Calculated conduction and valence band deformation potentials for 
different slab orientations (values in eV). The \emph{opposite} of the values in
the table can be regarded as bulk flexovoltage coefficients (in V). In both the (110)-
and (100)-oriented cases, the bending axis is assumed to be along [001], consistent 
with the experimental setup of Refs.~\onlinecite{pavlo} and~\onlinecite{Narvaez-15}. 
[SrTiO$_3$ is isotropic in the (111) plane, which means that the choice of the bending 
axis is irrelevant in this latter case.] 
\label{tab3}}
\end{table}

To illustrate the above ideas, it is an insightful excercise to compute the bulk ADPs 
under static slab bending by considering different surface orientations and mechanical boundary 
conditions. (The two extremes in the latter context are represented by the plate-bending
and beam-bending limits.) 
This is especially interesting in light of the experimental results of Ref.~\onlinecite{pavlo},
which concern the flexoelectric response (under bending) of SrTiO$_3$ slabs with 
different orientations.
Zubko {\em et al.}~\cite{pavlo} argued, based on their analysis, that the measured coefficients
can be interpreted reasonably well by assuming a purely bulk flexoelectric response, i.e. 
by neglecting possible surface effects.
It would then be desirable to compare the reported values with the existing 
theoretical estimates of the \emph{bulk} flexoelectric coefficients. 
Interestingly, reliable \emph{ab initio} calculations of these quantities have been 
recently performed~\cite{artcalc,Hong-13}, but the reported values
show a marked discrepancy with the experimental data: The first-principles
results for the bulk flexovoltage coefficients are of the order of $-10$ V,
and are systematically negative, while the measurements cluster~\cite{pavlo_review} around
$1-2$ V, with a positive or negative sign depending on sample orientation.
To account for such a discrepancy, the contribution of surfaces is typically 
invoked.~\cite{artcalc}
The concepts developed in this work, however, suggest that an alternative interpretation
is possible: The disagreement may be largely due to an unfortunate choice of the
energy reference when calculating the bulk contribution, and only to a lesser extent 
to the aforementioned surface effects.

Following up on this speculation, we report in Table~\ref{tab3} a complete overview
of the (bulk) conduction-band and valence-band ADPs in a SrTiO$_3$ slab subjected to
static bending.
(The three surface orientations and the choices of the bending axes are consistent 
with the experimental setup of Ref.~\onlinecite{pavlo}; for completeness we 
report the results for both the plate-bending and beam-bending regimes.)
Clearly, the calculated ADPs are much closer to the experimentally measured 
flexoelectric data of Zubko {\em et al.} than the bulk coefficients that 
were quoted in Refs.~\cite{artcalc,Hong-13}, corroborating our point.
(The values of Table~\ref{tab3} are of the order of 1 V; both
negative and positive values are present.)
Note that the difference between the plate-bending and beam-bending results 
is minor, which suggests that assuming one or the other regime might be
of relatively little importance for a qualitatively 
correct interpretation of the experiments.
Note also the relatively small orientation dependence, which amounts to about 1 V;
this supports the idea, recently proposed in Ref.~\onlinecite{Narvaez-15},
that an unusually large anisotropy (as measured by Narvaez {\em et al.} in
BaTiO$_3$ samples) is a clear signature of other effects (i.e. not bulk-like
in origin) being at play.

\section{Conclusions and outlook}

We have proposed a unified perspective on the theory of absolute deformation potentials
and flexoelectricity. To illustrate our ideas, we have used cubic SrTiO$_3$
(either in bulk or slab form) as a testcase.
The concepts developed here have important implications, both for the interpretation 
of the experimental measurements and for the macroscopic modeling of electromechanical
phenomena. 
In the latter context, our work highlights (and corroborates with quantitative
examples) both the arbitrariness of the bulk flexoelectric tensor, and the
need for a consistent treatment of bulk and surface effects in order to achieve
a sound physical picture. 
On a positive note, our work opens the way to the first-principles 
(and first-principles-based) study of an essentially unlimited range of 
phenomena related to inhomogeneous deformation fields, complementing long-range 
electrostatics with band-structure effects.

\section*{Acknowledgments}

This work was supported by MINECO-Spain (Grant No. 
FIS2013-48668-C2-2-P), and Generalitat de Catalunya (2014 SGR301).
We thankfully acknowledge the computer resources, 
technical expertise and assistance provided by the 
Supercomputing Center of Galicia (CESGA).

\bibliography{merged}

\end{document}